# Sabbath Day Home Automation:
## "It's Like Mixing Technology and Religion"

**Allison Woodruff**
Intel Research Berkeley
woodruff@acm.org

**Sally Augustin**
PlaceCoach, Inc.
sallyaugustin@yahoo.com

**Brooke Foucault**
Intel Digital Home Group
brooke.foucault@gmail.com

**ABSTRACT**
We present a qualitative study of 20 American Orthodox Jewish families' use of home automation for religious purposes. These lead users offer insight into real-life, long-term experience with home automation technologies. We discuss how automation was seen by participants to contribute to spiritual experience and how participants oriented to the use of automation as a religious custom. We also discuss the relationship of home automation to family life. We draw design implications for the broader population, including surrender of control as a design resource, home technologies that support long-term goals and lifestyle choices, and respite from technology.

**ACM Classification Keywords**
H5.m. Information interfaces and presentation (e.g. HCI): Miscellaneous.

**INTRODUCTION**

*It's Friday afternoon, and Trisha hurries to make a couple of last-minute adjustments to the X10 automation system in her home. The Sabbath is about to begin. During the Sabbath, which will last until Saturday evening, Trisha and her family will adhere to Jewish laws which prohibit them from manually turning electrical devices on or off. Instead of operating devices themselves, they will rely on the X10 system to control devices in the home. Trisha has been using X10 for this purpose since 1985, and her system has become quite sophisticated – for example, it will control kitchen appliances, it will open and close the skylight, and it will execute a complex sequence of on and off commands for lights throughout the home (even the one in the fish tank, as Trisha's husband believes the fish will only feed when the light is on). But the system is not there to side-step observance of an empty ritual. Trisha and her family aspire to spend the Sabbath in spiritual reflection. They feel the X10 system helps free them from mundane tasks and thoughts so they can focus on higher issues.*[1]

This paper reports the use of home automation by 20 families like Trisha's, households of Orthodox Jews[2] who have chosen to automate their homes for religious reasons.[3] The study has two primary motivations. First, it has been a rare opportunity to examine long-term user experiences with smart home technologies. The "smart home" [17] – a domestic environment capable of meeting and anticipating its occupants' needs using sensors and computational "intelligence" – is a core trope of ubiquitous computing research, but it is difficult to study. Smart home user experience research generally takes place in purpose-built laboratories, e.g. [19,21], which offer limited opportunities to study the interaction of technology with the complex and dynamic social processes of family life [11,17]. By contrast, many Orthodox Jewish families have been using home automation for decades; as we will see even their use of pre-computational technologies offers useful lessons because they presage the capabilities of more sophisticated smart homes. This community of lead users therefore offers valuable insight into real-life, long-term experiences with the use of home automation. Second, as Bell and others argue, the use of technology for spiritual purposes is a highly significant and understudied area [1,26]. The Orthodox Jewish community presents a wonderful opportunity for a case study in the use of technology for religious purposes.

---

[1] This scenario is based on discussions with a study participant. Participants' names as well as the name of the company referred to below as "RightSchedule" have been changed to protect anonymity.

[2] Jewish denominations vary by ritual and practice, with Orthodox (both Ultra- and Modern) being more strict than Conservative and Reform. Estimates vary, but of the world's Jewish population of 13 million, perhaps 2 million are Orthodox, including approximately one-half million in the United States [9]. Orthodox Jews as well as some Conservative Jews are observant of Sabbath laws. Some but not all of these use automation on the Sabbath (which is also referred to as Shabbos or Shabbat). While our sample was comprised primarily of Orthodox Jewish families, and these practices are most strongly associated with Orthodoxy, some of the families we visited were affiliated with less strict denominations.

[3] We understand some readers may wonder whether this is a religious "circumvention" and we discuss this further in the body of the paper.

While our investigation covered a wide range of topics, we focused in particular on questions such as the following: What is the relationship between home automation and religious practice? What is the relationship between home automation and family life? Accordingly, in this paper, we make the following novel contributions:

- We present a case study in the use of home automation technology by members of a religious community. In addition to describing specific practices, we discuss how a technology that performed mundane activities was seen to support spiritual experience and how participants oriented to the use of automation as a religious custom.
- We discuss the relationship of home automation to family life, arguing that in our participants' homes it played the role of an organizing system that revealed and reinforced the social order of the home [30].
- We draw on our in-depth examination of the Jewish perspective to inform and inspire design for the broader population, as the views of the families we studied provided a valuable counterpoint and prompted us to question common assumptions about smart home families and technologies.

The remainder of the paper is organized as follows. In the next section, we review related work. We then discuss our participants and method, findings, design implications, and conclusions.

**RELATED WORK**

There is an extensive literature on the domestic environment (see e.g. [24] for a survey), and [6] is an excellent example of how ethnographic findings in this environment may be applied to design. In addition, there is an abundant literature on smart home research; [2,17] provide an excellent introduction to this topic. However, a major limitation of research in this area is that the user experience has primarily been investigated in built laboratories [11,17], the notable exception being Mozer's intriguing experience as the sole occupant of the Adaptive House [25]. In this paper, we report a study of real-life, long-term user experiences with home automation systems that emerged in response to the needs of the residents.

Literature on spirituality and technology is significantly more sparse than that on smart homes, but compelling arguments have been made regarding its importance, and [1,26] call for more research in this area. The recent work of Wyche *et al.* on the use of communication technologies by pastors in megachurches is an excellent exploration in this vein [31], and we hope to further contribute to the literature on spirituality and technology by exploring the needs and experiences of the Orthodox Jewish population.

There is a rich and extensive literature on Jewish community and Jewish practice, including Sabbath rituals e.g. [8,9,15]. This research focuses primarily on issues such as the meaning, history, regulations, and experience of Jewish life. Many practical guides exist regarding Sabbath behavior, e.g. [5], and multiple professional and religious organizations provide guidance on and develop specialized Sabbath Day technologies for the Jewish population, e.g. [32]. Some mainstream commercial vendors also supply appliances such as ovens with Sabbath modes. The imagination of the press has occasionally been captured by these technologies [12,13]. These reports have focused on technology and rules for its use, rather than on user experience. To our knowledge this paper reports the first study of user experience with Sabbath technology – the first study to examine issues such as day-to-day practices with Sabbath technology, which technologies people choose to employ and find useful in their homes, user response to and perception of Sabbath technology, the interaction of Sabbath technology with family life, etc.

**PARTICIPANTS AND METHOD**

We recruited participants through advertisements in online forums, Jewish community email lists, and a Jewish newspaper, and through a developer of high-end automation systems for the Orthodox Jewish community. We visited a total of 20 homes and interviewed a total of 29 participants (16 men and 13 women) as well as having informal interactions with some additional household residents. Almost all participants were adults, at a variety of life stages (from young parent to retired empty-nester); many of the households had children of various ages. Participants had a range of occupations (e.g. dental hygienist, homemaker, CFO) and a range of roles in the synagogue (congregants, trainers-of-converts-to-Judaism, and rabbis). Thirteen of the homes were in the New York metropolitan area and seven were in the San Francisco metropolitan area. Homes were in both suburban and urban areas, and were predominantly single-family detached homes.

Most households were Modern Orthodox although some were Ultra-Orthodox. A small number were Conservative households that adhered to fairly strict practices. In general, participants were fairly highly integrated in mainstream American culture and product consumption, consistent with Diamond's discussion of Modern Orthodoxy as "a style of Jewish life that blends the rituals of Orthodoxy with the rituals of the mall, the gym, and the Long Island Rail Road" [Gross, quoted in 8]. All participants used some type of automation to adhere to the Sabbath code of conduct outlined in Jewish law, ranging from quite basic to very sophisticated. Quite a few of the participants, including some who had become Orthodox later in life, had decades of experience using automation on the Sabbath (sometimes dating back to childhood), either in the same home or in multiple homes.

In addition to the participants we visited in their homes, we interviewed the developer of a high-end automation system, and had additional conversations with many other members of the community by telephone, over email, and in person. We also visited Judaica stores and Jewish neighborhoods and had an overnight Sabbath visit with one of the families.

We conducted home visits in December 2005 and May 2006. Visits typically lasted one-and-a-half to two hours, and consisted of a semi-structured interview, a tour of relevant parts of the home including demonstrations of technology in place, and a photo-elicitation exercise designed to prompt discussion [18]. All interviews were video-taped or audio-taped. All interviews were transcribed verbatim, resulting in a corpus of approximately 1300 pages (approximately 340,000 words). We performed an affinity clustering of the data to identify emergent themes [4]. Note that our analysis has the attendant advantages and disadvantages of an "outsider's" perspective, as none of the members of our research team is an Orthodox Jew.

**THE SABBATH EXPERIENCE**
In this section, we discuss the temporal structure of the week, prescribed and proscribed activities, orientation to technology and information, and respite and renewal.[4]

**The Structure of the Week**
The Jewish belief is that the ultimate aim on Earth is to transform Earth into God's dwelling – this is the work six days of the week [5]. Participants characterized this time as busy or hectic, a time of changing things and manipulating the world. During the week, in addition to conducting everyday activities, one prepares for the Sabbath, particularly on Friday afternoon as the Sabbath approaches. Preparing for the Sabbath includes preparing food, preparing the physical self (bathing, cutting nails, cutting hair), and preparing the domestic environment (getting fresh cut flowers, setting the lights and automation).

A few minutes before sunset on Friday, the Sabbath begins and continues until after sunset on Saturday. On the Sabbath, the aim is to experience a taste of Paradise, the world to come. Although it is often colloquially referred to as a day on which to "rest," literally it is a day on which to "cease" creating, a serene period during which one strives to have minimal impact on the world and to leave the world unchanged. During the Sabbath, the Orthodox Jew submits to external processes and "renounces his autonomy and affirms God's dominion over him" [Tsevat, quoted in 9]. By surrendering control, Orthodox Jews strive to clear their minds to focus and reflect on larger issues. The Sabbath is described as a time of peace, relaxation, and reflection.

> Isaac: "There's clearly preparation that go on ahead of time. But pretty much when the Sabbath begins, life comes to a screeching halt… from a tornado to complete calm. And uh you spend the next 25 hours within that envelope of calm, sort of disconnected from the real world, if you would."

> Lisa: "On the Sabbath, among the things that we are conscious of not doing is changing things."

> Rachel: "It's like time standing still."

**Prescribed and Proscribed Activities**
A number of activities are explicitly encouraged on the Sabbath – going to synagogue, spending time with family and friends, studying religious materials, reflecting, taking naps, and going for walks. The activities are such that the Sabbath is a "beautiful" day, with many pleasant social and sensory experiences such as nice scents and delicious food.

Activities associated with work or creation are explicitly prohibited on the Sabbath. Specifically, the *halacha* (the collective corpus of Jewish religious law) forbids 39 activities that were associated with maintaining the temple in ancient times, such as sowing, baking, or kindling a fire. While the halacha itself is from ancient times, it has been continuously interpreted and reinterpreted by rabbinical authorities as society and technology have evolved. In these interpretations, a wide range of activities related to secular life are forbidden, from carrying a key to synagogue to reading a newspaper. For our purposes, one of the most significant modern interpretations is that it is forbidden to turn electrical devices on or off. (There are several different arguments as to *why* this might be prohibited, e.g. that turning electrical devices on and off is associated with kindling or extinguishing a fire, but in any case there seems to be consensus that it *is* in fact prohibited.) The objective of these prohibitions is not punishment or sacrifice *per se*; rather their aim is to create an appropriate environment and mindset in which to receive the Sabbath.

> Isaac: "On a day when all standard activity stops and work stops, et cetera, you end up with a much clearer lens into any relationship you have with God."

> Luke: "We imitate what God did… for six days we exert mastery over the physical world. We do things. We create. We write. We plant and sow and reap and cook… just as God himself set the whole thing up by creating the world. Then on the seventh day, just as he rested… similarly the underlying principle of Shabbat… is to refrain from creative activity… and in particular to refrain from demonstrations of mastery over God's world. This is a day of – to focus on the spiritual rather than the physical. And so we refrain. Ideally when we leave Shabbat 25 hours after we enter it, we will have made – we will have had as light a footprint as possible on the earth… I think you can see how turning a light switch on and off, though it involves no work, is nonetheless very much a demonstration of mastery."

**Technology and Information**
Technology and information were often directly associated with stress or worry, and respite from them was an important part of the Sabbath experience. (Note that these issues arose spontaneously and not in response to a question about technology; many of the other issues we report arose spontaneously as well.)

---
[4] Our findings resonate with reports of Jewish practice in other literatures, e.g. [9,15]. However, note that there may be alternative viewpoints to those expressed by our participants. While our findings regarding the Sabbath experience provide valuable context, and we make a contribution by interpreting them from the HCI perspective, our primary novelty lies in our findings regarding Sabbath Day home automation reported in the subsequent sections.

Leah: "Do I want to know about the news? Do I want to receive telephone calls – be able to hear them? I personally shut them out. A lot of people leave on their answering machine and so they can hear it. I personally find that invasive on the Sabbath… Do I want to know the news? Probably not. It could only be bad and there's a sense that you don't want something invasive coming into your Sabbath celebration. The bad news will wait, and it does."

Stuart: "Not driving a car; not answering the phone; not turning –"
Lisa: "– not cooking –"
Stuart: "– on the computer. These are liberating things and this is part of the Sabbath."

Leah: "It's *muktzah*… something that you don't touch on the Sabbath because it's something that's essentially forbidden… If a kid comes up and starts fooling with the laptop you say to him, 'It's *muktzah*.' It's not something that is within the realm of the Sabbath."

Further, technology was often positioned as being at odds with a focus on family and friends (in fact, one family joked that they think of themselves as "Post Modern Orthodox"), and time away from technology was seen as nurturing relationships.

Lisa: "It's a sense of being together and caring for each other and loving in low tech."

Luke: "A lot of the times during the week Talia will say, I was really looking forward to having you at home. And so I get home and she says, great, now we can knock [off] some of the TV we've accumulated. On Shabbos when I'm home we really talk… the point is it's something that forces us to deal with ourselves as humans and as the primary focus rather than we'll watch this TV together and that's sharing."

**Respite and Renewal**
Overall, the Sabbath is a time of renewal, consistent with a basic human desire to be refreshed psychologically. It encompasses activities such as reflection that are restorative [20], and provides respite from weekday activities, stressful thoughts, technology, and information.

Talia: "There's just something about it that I feel more peaceful on Shabbat and it's like a recharging of my batteries every week."

Lisa: "We use a lot more technology during the week because it's interactive and we're changing things… the Sabbath is in some ways like taking a vacation from the week."

Hilary: "[The Sabbath is] a time-out from the 'hectic-ness' of every day life, to stop and refocus, and relax and regain your inner balance."

**HOME AUTOMATION SYSTEMS FOR THE SABBATH**
As mentioned above, manually controlling electrical devices is generally prohibited on the Sabbath. However, many Orthodox Jews use timers or other automation technologies on the Sabbath. The reasoning is that, although an Orthodox Jew should not do anything *during* the Sabbath that has impact on electrical devices, they can perform acts *in advance* of the Sabbath that will have impact during the Sabbath. By this reasoning, many rabbis teach that timers are acceptable on the Sabbath but motion sensors are not because if one triggered a motion sensor, one would be affecting the state of the world directly with one's body (accordingly, devices such as motion sensors must be automatically or manually disabled on the Sabbath). We discuss these issues further below.

**Types of Automation**
In this section, we discuss the types of automation used in the homes we visited. There was a wide range of capabilities, from scheduling events at fixed times to using environmental sensors such as a rain sensor to trigger the closing of a skylight. The homes we visited fell into three categories, based on the technology in use: Timers (9 homes), X10 (6 homes), and RightSchedule (5 homes). We do not have data on overall trends and do not suggest that this distribution is representative of the overall population.

*Timers*. In the Timer homes, automation generally consisted of a handful of rotary timers or digital timers (approximately two to six would be typical), most often for places or devices with relatively predictable use patterns, such as the dining room chandelier. Commercial appliances were also often useful on the Sabbath, e.g. TiVo to record programs to watch at a later time or ovens with a Sabbath Day mode to disable automatic shut-off.

A number of "low-tech" solutions typically complemented the use of automation in these homes. Lights in areas of the home with less predictable use, such as bathrooms, were often simply turned on before the Sabbath and left on for the duration. Kitchens also had a number of additional appliances such as "blechs" (specialized hot plates that adhere to halachic constraints regarding food preparation).

Abigail: "Then there's the kitchen, which is a timer fest in and of itself."

Lisa: "The refrigerator light has been turned off since the day it arrived… we just took the bulb out so that we don't have it turning on and off on the Sabbath [when we open the refrigerator door] and don't have to worry about it... But that's not exactly automated. That's- that's anti-automated."

*X10*. Several of the homes used X10, a system for using household wiring to send digital data between devices. X10 users would set up a schedule on a computer, which would then send commands to various X10 controllers throughout the home to turn devices on and off. X10 users tended to develop fairly complex lighting schedules that covered more areas in the home than timers, and X10 was used to control a wide range of devices from lights to towel warmers to skylights. X10 is notoriously unreliable and tricky, so the people using this system tended to be fairly tech-savvy "do-it-yourself"ers who had impressive stories of both successes and failures. In addition to using X10, these homes usually employed a few "low-tech" solutions, such as "blechs."

*RightSchedule*. Several of the homes we visited used a high-end system developed specifically for the Orthodox Jewish community. The system includes a rule-based program that uses the Jewish calendar and the families' specified preferences to dynamically generate a schedule that interfaces with controllers (typically X10) installed by the system developer, as well as with other automation systems in the home, such as lighting, sprinkler, security, entertainment, and security systems. RightSchedule knows the Jewish calendar for the next several decades, and takes into account the varying requirements for different holidays.[5] For example, the dairy oven may be turned on by the automation system only once a year, for the Shavuot holiday.

RightSchedule is an expensive system, costing tens of thousands of dollars or more. The system is targeted at the control of numerous devices in large, luxury homes. For example, one such home that we visited had eleven bedrooms and nine bathrooms. On a three-day holiday, approximately 5500 on/off signals are sent to the 175 devices in this home, including not only myriad lights and kitchen appliances, but also more exotic items such as an elevator, a large bank of drapes, and a roof that slides back to create the religiously required open air "booth" during the annual Sukkot holiday. In many homes, RightSchedule also blinks the lights to provide reminders, for example that services at the nearby synagogue are about to begin.

All the clients we met had established a highly personal relationship with the developer, Mr. Herschel. He is a home automation coach of sorts, a key influence in how his clients think about automation and the strategies they use. He meets with clients for extensive planning sessions before their systems are installed and becomes intimately familiar with their lives. His relationships with them are similar to those that a physician might establish (as opposed to those that an electrician or a plumber might establish). He is treated as a professional and a peer by his high-end clientele, frequently staying in their homes when traveling. Participants reported that Mr. Herschel is very personally committed to the system, and by all accounts he and his staff provide exceptional service. Mr. Herschel provides a buffer between his clients and the implementation. Debugging is virtually always done by Mr. Herschel and his staff rather than by household occupants, and in fact some clients seemed largely unaware of how the technology worked or even that X10 was present in their homes. Clients routinely orient to Mr. Herschel as the interface to the system, often phoning or emailing him to request changes to the system rather than using the computer interface to make changes themselves – even the simplest changes, such as making a light stay on half an hour later or changing the house from "vacation" mode to "here" mode (which can be done just by pressing a button on the wall), were likely to be made through Mr. Herschel or his staff.

**Motivation and Use on Non-Religious Days**
In almost all households, participants expressed that their primary motivation for getting and continuing to use automation was religious observance, and occupants were generally quite satisfied with their systems. Most expressed that it would not be worth installing and maintaining automation if it were not for the Sabbath. However, once it was in place, there was some use of it on other days.

> Trisha: "Once we had the system in then we do other things… once it's here and we spent the money on the switches, then we have them go on, you know, go off at night..."
>
> Noam: "Well, it all started from the Sabbath part. From there it… went to the entire week."

One might imagine that automation might be co-opted for significant non-Sabbath use that might even eventually overshadow Sabbath use, but this did not generally seem to have occurred. Almost all participants reported the systems would not be worth getting if not for religious purposes, and utility on non-religious days tended to be limited. While practices varied, the most useful non-religious functions appeared to be: (1) turning lights on and off automatically when no one was home, to make the house look "lived in" for security; and (2) turning off lights that were out of sight and/or out of reach, either automatically or with a remote at the side of the parents' bed.

### AUTOMATION AND THE GOALS OF THE SABBATH
Interestingly, while information and communication technologies and interactive technologies were perceived as being at odds with the Sabbath, automation technology was perceived as promoting the goals of the Sabbath. In this section, we discuss three primary ways in which automation was seen to contribute to the goals of the Sabbath.

**Relief from Burden**
Recall that on the Sabbath one strives to clear one's mind and remain in a reflective state. Ceding mundane responsibilities to the automation system "freed up" the minds of participants so they could focus on issues that transcended daily life. Ironically, technology was used to free the residents from technology; more technology provided the illusion of less technology.

> Rachel: "It really is liberating to just know the kind of like the support mechanisms are taken care of and you can just go ahead and live and function and meet your goals. And the system's kind of there to support you. It's a nice feeling."
>
> Rachel: "We try, as a couple, to work on having those things in place so that we can really enjoy our Sabbath, our Shabbat,

---

[5] In addition to observing the Sabbath, the Jewish community also has a complex holiday schedule. Holidays have many similarities with the Sabbath, as well as special requirements of their own. Automation is therefore useful on holidays as well as on the Sabbath, and when holidays fall adjacent to the Sabbath, the Jewish community needs to plan a schedule (including automation) for two or even three consecutive days. Our observations regarding use of automation on the Sabbath are usually germane to use of automation on Jewish holidays as well.

with our family and friends and not have to worry about – when you have all the details in place, you can get down to business and look at the bigger picture."

Hilary: "Having our home automated means that we're not doing the things that we normally do on the other six days of the week. These things are taken care of for us; I'm not cooking; I'm not washing…"

Robert: "Once it's done… we ignore it… we've set up and we're done …we don't give it any thought."

Nathan: "I can avoid all the electronics; thinking about them, or why they're on or why they're off."

Joseph: "It's a simplified life. We don't have to deal with the lights. We don't have to deal with, but it just happens. And it just happens without us, and we benefit from it. And that's really the power of it uh from our perspective."

**Enhancement of the Experience**
There was a sense that the automation concealed what was going on behind the scenes, giving a pleasing final effect.

Joseph: "The notion of going on with your life with the simplicity of the Sabbath while all this complex stuff is happening elsewhere."

Ephraim: "You don't see [the wiring and stuff like that] but then you see a beautiful end product."

The mechanical functions the automation system performed were not considered central to the Sabbath experience and, while valuable, automation was rarely characterized as necessary. Participants explained that most Sabbath activities would occur with or without automation – simply leaving lights on and appliances running is a viable alternative (albeit one that is "wasteful" of energy). The impact of automation on the Sabbath is therefore fairly subtle. For example, when non-essential devices were automated, people were more likely to use them on the Sabbath – a family with automation might have wall sconces turn on for part of the Sabbath, while without automation they might rely on only the main overhead light in the room. Technology was typically characterized as "enhancing" the Sabbath experience, adding to the atmosphere and the ambiance (e.g. prettier lights, running water in a fountain, better food, etc.). The Sabbath is a sensory experience, and the automation complements this.

Lisa: "It just makes it more… physically comfortable… but, most of what happens, would happen anyway."

**Surrender of Control**
The Sabbath is associated with submission to external processes and entities. Our participants' narratives regarding control were complex. While there were some situations in which automation was associated with a sense of being in control, a more prominent and fascinating theme that emerged was the benefit of intentional surrender of control on the Sabbath. There is value in the *inability* to control – if a process can not be overridden or controlled in the moment, there is less motivation to actively attend to it, and passivity can facilitate reflection. As used, the automation systems in place are therefore very consistent with and reinforce the Sabbath experience. Accordingly, participants simply experienced the effects of the automation system as they occurred. In fact, as in other areas of their lives, tolerance for undesirable outcomes appeared to be part of the value system. The systems had generally evolved to the point where they were fairly reliable, but errors did occur. When the automation system did something inconvenient, the ideal attitude appeared to be calm acceptance rather than frustration.

Nathan: "God controls everything – that's the way we live it… I do it [set up the lighting] and if it works it works, and if it doesn't that's what God wants it."

Reuben: "I don't get upset if it doesn't happen. But it's nice when it does. The fact that everything works and is in sync, and you know, and everything runs smoothly. We like when our lives are running smoothly."

In fact, setting up and then relinquishing control to the automation system was even seen as valuable training in the principles of the Sabbath.

Leah: "You're used to doing things at will… And for the Sabbath, particularly, you preprogram and you're done... as soon as the Sabbath comes in, I'm out of control. And that – in some ways that's a very interesting lesson in life, and it's one of the lessons of the Sabbath... That could be very frustrating, and it's also a good lesson in life. You have to live with what you've got. So, that's the part of your automation that it's a fixed schedule and you need to live with it."

Leah: "You have to get used to restraining yourself from operating or being comfortable with not operating. I mean this is a very difficult thing and it takes a while to get used to, if you haven't been brought up that way. Which I was not."

**USE OF AUTOMATION AS RELIGIOUS CUSTOM**
Use of home automation appeared to have become a religious custom for the participants, both a ritual in the home and a sign of affiliation with the community. Over the years, automation and/or the atmosphere it created became associated with religious commitment and ritual.

Stuart: "I would certainly want… to keep the same pattern.… It's a shared [commitment between me and my wife] – how we run our home and our commitment to the Sabbath."

Reuben: "Also I kind of like the fact, it feels like it lends to the Sabbath feeling on some level. When the lights go off, it does add… it's not a major part of the feeling… a lot of it's intangible things… but it does add to the atmosphere… It's probably more of a nurture thing, not nature so to speak. Because I grew up with it. It's just what you're used to during the Sabbath, the lights kind of go off on a timer."

Stuart, who had lived in his home with a timer system for over thirty years, spoke of the significance of maintaining the familiar experience of the automation system, and explained that he would be reluctant to make any changes, even to get increased functionality.

Stuart: "I could imagine some of the things that might be available, that we just wouldn't be interested in, necessarily,

because we want to retain the sanctity of the day and in kind of in the way that we've known it for all these years."

The skeptically inclined may wonder whether the use of automation on the Sabbath is a "circumvention," a "subterfuge," or even a "cheat." This is a complex question. Religion is the fundamental organizing principle in many of our participants' lives. Participants evidenced a strong commitment to the Sabbath, and the use of automation notwithstanding, the Sabbath as a lived experience was dramatically different from the quotidian routine of the rest of the week. The appropriateness of automation is a subject of debate in rabbinical interpretation, and multiple institutes specialize in halacha and technology, e.g. [32]. Many members of the Orthodox Jewish population enjoy debating and bonding over topics ranging from technology to kosher "fake bacon" [8,9].

Luke: "We often have a great deal of fun on Shabbat when we aren't sure about doing something or when we have a difference of opinion, pulling down our references off the book shelves and going through the legal codes and debating it. That's just a fun activity, you know, in its own right."

While opinions vary, especially regarding particulars, a key point is that automation is an external process. The use of automation is therefore consistent with other notions of the Sabbath; just as one prepares food before the Sabbath for consumption on the Sabbath, one can prepare an orchestrated lighting experience in advance of the Sabbath.

Luke: "My own analogy… is that when God rested on Shabbat, the sun did not freeze in place and wait for 25 hours. And the flowers did not, you know – the sun rose and the flowers in response opened, and they all functioned as they were predetermined to function and continue on Shabbat. And similarly I feel that the use of timers is exactly the same thing. Prior to Shabbat I set the rules, and the rules now flow in motion throughout Shabbat."

Nathan: "[The water urn] turns itself on and off; when it gets too cold, it turns itself on; you can see it there. It boils on its own accord. It doesn't have to keep the Sabbath, we do."

While a debate of the halachic correctness of automation is beyond the scope of this paper, in any case complicated strategies for dealing with forbidden activities have long been part of Jewish custom and Jewish identity and certainly predate the existence of modern-day technology [9]. For example, the Jewish community has a very old practice of having human servants perform forbidden activities during the Sabbath, and Dundes argues that automation is a modern-day extension of this practice [9]. Consistent with Dundes' suggestion that "circumventions" can legitimately be considered customs and are an adhesive basis for Jewish identity, use of automation appeared to be a unifying force that participants associated with commitment to the Sabbath and Judaism.

Luke: "I think Shabbat more than the rest of the time I really feel distinct as a Jew from my surrounding western culture because nobody else does this stuff with the lights, you know."

Rachel: "We're so blessed to live in a modern age where we can have your cake and eat it, too. You can have a light on, a lamp on, or turn a Shabbat light on and be able to see what you're doing. And still show respect to religion and tradition and be able to have a living Judaism that we can pass on to our children. They can see how we observe it… it's like mixing technology and religion. It's just cool."

## HOME AUTOMATION AND FAMILY LIFE

The automation system became interwoven with the family lives of our participants in a number of ways. Naturally, family life is strongly connected to the religious practices we have discussed in previous sections, and automation is one mechanism that integrates family life and religious life.

Polly: "All the little bits and pieces [of religious practice and religious technology] that we put together and that we decided over the years would be part of our family life."

Leah: "People made the transition to greater and greater observance, and accommodated their homes and their social relationships accordingly."

More broadly, the home automation system reflected and shaped the routines, expected behaviors, and social relations of family life – the social order of the home. In this way, it was similar to the organizing systems discussed by Taylor and Swan [30]. Taylor and Swan report on paper artifacts such as calendars and todo lists, and we extend their analysis by observing that a technological artifact (the home automation system) can be appropriated by household occupants to reflect and reinforce the social order of the home. Home automation systems are particularly rich organizing systems because they can act "autonomously" to modify the physical environment, and because they are embodied in the objects and infrastructure of the home. In the remainder of this section, we focus on two illustrative examples of the interaction of automation with social order: automation as a resource for influencing behavior, and participants' interpretation of the role of automation.

### Automation as a Resource for Influencing Behavior

Automation provides cues as to what actions are expected of household occupants at what times, and it is therefore a resource for influencing behavior – for example, when kids are in the recreation room late at night, a light turning off sends a "message" that it is time to go to bed. Further, actions taken by the automation system are a useful resource for facework, e.g. lights that turn off are a resource for suggesting that it is time for guests to leave after a Sabbath meal (they are not however a mandate; guests do sometimes stay long after the dining room light goes out).

Daniel: "Let's say at about 10:00 o'clock the lights would start to dim, 10:15, a little bit more. By 11:00 o'clock they would get the message… You can't tell guests, you know, I think it's time to go. But like this –"

Joseph: "The story was told to us… The lights would go off at 3:15 in the afternoon in the dining room… if you were not good company, rabbi would say, 'Oh, the lights are off. I guess it's time to finish the meal and say goodbye.' But if they were good company, the rabbi would say, 'Wait 15 minutes, they're going back on.' [laughter] Now, this may be apocryphal, I don't know. But it's a great example of what you can do with these things."

There are also subtle cases in which the automation suggests routine, even when it is not intentionally designed to do so. Daniel discusses how the lights gently guide his behavior and create an environment that is conducive to religious observance of fasting.

Daniel: "We have a holiday called Yom Kippur where we don't eat… When we come home from the synagogue, basically we go to sleep because there's no – you can't eat… So, I always find it unique. I come into the home. All the lights are on except the dining room light, because that light goes out while we're in synagogue because it reminds you it's Yom Kippur."

It is important to clarify that while automation naturally interacts with schedule, there was little indication that participants led highly choreographed lives or were slaves to a schedule imposed by automation on the Sabbath. Participants employed a number of strategies (e.g. leaving generous time buffers in automation schedules) such that automation encoded and subtly influenced the routines in the home, rather than regimenting them, and the technology was generally thought to "work with your lifestyle."

**Interpretation of the Role of the Automation System**
Interpretations of the automation system and its actions were varied and complex. However, it was quite common for participants to attribute meaning to actions taken by the automation system, and sometimes to associate them with expected behavior. First, participants sometimes oriented to the automation system as an extension or proxy of the person setting the schedule of the automation system. A fascinating example is that many of Mr. Herschel's clients oriented to the system as an extension of him.

Aleksei: "We usually call it the Herschel System… Well, the truth is, I think of it as Mr. Herschel himself…"

Through the automation system, Mr. Herschel becomes highly present in the home, and actions taken by the system may even be interpreted as him making comments on the behavior of the occupants.

Aleksei: "Let's say I would be working out late… and then all of a sudden all the lights go out… But then again, he [Mr. Herschel] is probably right. I shouldn't be working out that late anyway."

Second, automation was strongly associated with caretaking, anticipation, and guidance – roles[6] such as servant [16] (sometimes quite a wise servant), mother, and wife. There were also occasional allusions to more godlike or omniscient characteristics.

Rachel: "Nurturing like a parent or a god… Always thinks to take care of the needs… Like Mother Earth taking care of everybody's needs and foreseeing the needs and planning ahead, very mothering, nurturing."

## IMPLICATIONS FOR DESIGN

The experiences of our participants suggest a wide range of design possibilities. One of the themes that we found most striking was the orientation to external forces – external mandates, processes, community, etc. This perspective is in stark contrast with traditional visions of the smart home, which focus on control and mastery. For example, Hamill argues that "smart domestic devices should put people firmly in control" [16] (see [7] for a review of similar arguments), and Davidoff *et al.* argue that smart home technology should be used to give families "more control of their lives" [7].

While compelling arguments have been made that smart home design should focus on domesticity and cultural context and variation rather than technology *per se* [2], to date it seems that smart home families are somewhat homogenously envisioned as control-oriented and goal-oriented, with a focus on the achievements of the family and its members. All families are not the same, and the families in our study provide a useful counterpoint. They aspired to surrender control to God and external forces, particularly on the Sabbath. Rather than seeking to exercise dominion, they sought to have as small a footprint on the earth as possible. They sought to explore harmonious connections with God, the broader Jewish community, and the natural world. We believe this alternative perspective inspires new investigations in the design space, and we discuss three specific areas of interest below.

**Surrender of Control as a Design Resource**
As mentioned above, traditional wisdom argues for a high degree of end-user control. Our findings prompt us to consider a richer set of options in the design space. Studies suggest that there are situations in which surrendering control offers significant psychological benefit. Multifaceted drive theory proposes a number of basic human drives such as autonomy and power. Reiss argues that individuals are motivated to aim for moderation in the satiation of these drives; for example, "When a person experiences more power than he or she desires, the individual is motivated to be submissive for a period of time to balance experience toward the desired rate. When a person experiences less power than he or she desires, the individual is motivated to be domineering for a time" [27]. This would suggest that some individuals might be better served by systems that present a wider range of options for autonomy, e.g. that certain individuals at certain times would benefit from experiences that give them a sense of another entity being in control. Further, Reiss also presents findings that religiosity is associated with a desire for low autonomy [27]. In combination with the experiences reported by our participants, this would suggest that designs for spirituality might explore a range of experiences related to the surrender of control.

---

[6] These points arose largely in response to a prompt about "personality traits" of the system; we are not suggesting that participants think of their automation systems as people, but rather that we found it informative to consider the traits and roles they associated with the system when asked. For example, one might have imagined that automation might be conceived of as tyrannical or incorrigible, but this did seem to be the case.

Plainly, we are not arguing that people should not have any control of their devices. We are however proposing that giving up control can be beneficial or desirable in some situations, and that this is an interesting design space to explore. Surrender of control has been examined in, for example, game design in an urban environment [3]; here we propose that it is valuable to explore in domestic settings, as well as for spiritual purposes. In some ways it may seem ludicrous to propose that something so banal as an automation system might be a resource for balancing basic human drives relating to autonomy and power, or for facilitating spiritual experiences. One might imagine for example that such needs are more appropriately met through social interaction or visits to magnificent churches or awe-inspiring natural vistas. Nonetheless, the experiences reported by our participants suggest that everyday objects can in fact be appropriated for these purposes.

**Support for Lifestyles and Long-term Goals**
Over time, automation became interleaved with the religious practices and the family lives of our participants, both reflecting and reinforcing the participants' long-term goals. Specific designs and artifacts are understood to guide users towards certain behaviors, e.g. [22,14], and people are known to try to create environments that are conducive to their goals [23]. This suggests that, in addition to offering benefits that have been traditionally proposed such as convenience and efficiency, domestic technologies are resources that can be used to facilitate chosen lifestyles. Davidoff *et al*. [7] explore how technology may support one such lifestyle – the lifestyle of the busy family [29], with its attendant value system. A wide range of other lifestyles exist, with different value systems, and it is interesting to consider how technology physically embodied in the home might support lifestyles such as green living, slow living, spirituality, etc.

**Respite as a Mandate and as a Community Experience**
As the vision of "anywhere, anytime" access to technology becomes increasingly real, concerns mount regarding a host of issues such as technology addiction, work-life (im)balance, social isolation, cognitive overload, etc. Understanding these issues and proposing resources to allow people to negotiate the boundaries between technology and other parts of their lives is a central challenge for HCI research. Our participants offer one highly informative example of how a community orients to and sets boundaries with technology and information, suggesting the following implications for the design of respite for the broader population: (1) externally mandated respite for strictly bounded periods of time is one viable model and (2) respite designed at the level of the community may be more effective than individual respite.

Consider the story of how following religious law helped Talia break a technological addiction.

> Luke: "If I can tell a Talia story. When the twin towers thing happened… she turned into a complete news junky for that. And she literally stopped going to work and just watched, you know, CNN and all those networks finding out everything that was going on until as it happened Rosh Hashanah came, which was a holiday. It came before Shabbat. And she had to stop because that was the law. And by the end of those two days of being turned off the media, she was still, you know, concerned and she still followed things up."
> Talia: "But I was functional again. I went back to work the next day."
> Luke: "Right, she was no longer obsessed…"
> Talia: "I lived with the TV. I didn't go to work. I didn't even call to tell them I wasn't coming in. I just shut off from the world… And that break meant, you know, so much. So, I mean it really, it's very helpful to have to turn the world off."

The Sabbath is an external mandate, a mechanism that "forces" desirable behavior. This external mandate is likely far more effective than having individuals set internal goals, and seems to be most effective when reinforced by the community (see also the argument in [23] that lifestyle change usually takes large-scale effort). We were told that it is extremely difficult to observe the Sabbath when one is isolated from the community and when one is surrounded by other people who are *not* observing the Sabbath. By contrast, being with members of the community makes it easier to be observant, because the surrounding temporal rhythms and institutional support align with one's goals [28]. For example, some apartment buildings in Israel have central timing and wiring systems to control lights and other devices throughout the building on the Sabbath.

It is natural to conceive of sanctuary as an individual and isolated experience, and designs for digital shelter have tended to focus on individuals or households, e.g. [10]. However, the experiences reported by our participants suggests that objects or practices at the level of individuals or even households are not nearly as likely to be effective at facilitating respite as designs targeted at communities. Perhaps respite from technology should be reconceived as a community experience.

**CONCLUSION**
In this paper, we have presented a study of 20 American Orthodox Jewish families' use of home automation for religious purposes. We have explored the relationship between home automation and religious practice in this community, discussing how a technology that performed mundane activities was seen to support spiritual experience and how participants oriented to the use of automation as a religious custom. We have also discussed the relationship of home automation to family life, arguing that in our participants' homes it played the role of an organizing system that revealed and reinforced the social order of the home [30]. We have further drawn design implications for the broader population, including surrender of control as a design resource, support for long-term goals and lifestyles, and respite from technology. In future research, we intend to investigate the use of domestic technologies by other populations influenced by clearly articulated value systems, such as the use of technology in the service of green living.

**ACKNOWLEDGMENTS**
We gratefully acknowledge our participants for their kindness in welcoming us into their homes. We thank Paul Aoki, Scott Mainwaring, Victor Buchli, Ryan Aipperspach, Tom Igoe, Sue Faulkner, and Francoise Bourdonnec for valuable discussions, comments, and support.